\documentclass[sigconf]{acmart}

\usepackage{color}
\usepackage{multirow}
\usepackage{amsfonts} 
\usepackage{subfigure}
\usepackage[ruled,vlined]{algorithm2e} 
\usepackage{hyperref}

\newcommand{\red}[1]{#1}
\newcommand{\blue}[1]{#1}

\AtBeginDocument{%
  \providecommand\BibTeX{{%
    \normalfont B\kern-0.5em{\scshape i\kern-0.25em b}\kern-0.8em\TeX}}}

\setcopyright{iw3c2w3}
\copyrightyear{2021}
\acmYear{2021}
\acmDOI{10.1145/3442381.3450006}

\acmConference[WWW '21]{Proceedings of the Web Conference 2021}{April 19--23, 2021}{Ljubljana, Slovenia}
\acmBooktitle{Proceedings of the Web Conference 2021 (WWW '21), April 19--23, 2021,
Ljubljana, Slovenia}
\acmPrice{}
\acmISBN{978-1-4503-8312-7/21/04}

\begin{document}

\title{Meta-HAR: Federated Representation Learning for Human Activity Recognition}

\author{Chenglin Li}
\affiliation{%
  \institution{University of Alberta}
  \city{Edmonton}
  \country{Canada}
  \postcode{T6G-2R3}
}
\email{ch11@ualberta.ca}

\author{Di Niu}
\affiliation{
  \institution{University of Alberta}
  \city{Edmonton}
  \country{Canada}
  \postcode{T6G-2R3}
}
\email{dniu@ualberta.ca}

\author{Bei Jiang}
\affiliation{
  \institution{University of Alberta}
  \city{Edmonton}
  \country{Canada}
  \postcode{T6G-2R3}
}
\email{bei1@ualberta.ca}

\author{Xiao Zuo}
\affiliation{
  \institution{Tencent}
  \city{Shenzhen}
  \country{China}
}
\email{royzuo@tencent.com}

\author{Jianming Yang}
\affiliation{
  \institution{Tencent}
  \city{Shenzhen}
  \country{China}
}
\email{kimmyyang@tencent.com}


\begin{abstract}
\label{sec:abstract}
Human activity recognition (HAR) based on mobile sensors plays an important role in ubiquitous computing. However, the rise of data regulatory constraints precludes collecting private and labeled signal data from personal devices at scale. Thanks to the growth of computational power on mobile devices, federated learning has emerged as a decentralized alternative solution to model training, which iteratively aggregates locally updated models into a shared global model, therefore being able to leverage decentralized, private data without central collection. However, the effectiveness of federated learning for HAR is affected by the fact that each user has different activity types and even a different signal distribution for the same activity type. Furthermore, it is uncertain if a single global model trained can generalize well to individual users or new users with heterogeneous data. In this paper, we propose Meta-HAR, a \emph{federated representation learning} framework, in which a signal embedding network is \emph{meta-learned} in a federated manner, while the learned signal representations are further fed into a personalized classification network at each user for activity prediction. In order to boost the representation ability of the embedding network, we treat the HAR problem at each user as a different task and train the shared embedding network through a Model-Agnostic Meta-learning framework, such that the embedding network can generalize to any individual user. Personalization is further achieved on top of the robustly learned representations in an adaptation procedure.
We conducted extensive experiments based on two publicly available HAR datasets as well as a newly created HAR dataset.
Results verify that Meta-HAR is effective at maintaining high test accuracies for individual users, including new users, and significantly outperforms several baselines, including \emph{Federated Averaging}, \emph{Reptile} and even centralized learning in certain cases. Our collected dataset will be open-sourced to facilitate future development in the field of sensor-based human activity recognition.
\end{abstract}



\maketitle

\section{Introduction}
\label{sec:intro}

Human activity recognition (HAR) is the problem of recognizing human activity types based on mobile sensor data, playing an important role in ubiquitous and pervasive computing. State-of-the-art approaches rely on deep neural network models to replace traditional manual feature engineering and have greatly improved the accuracy of HAR~\cite{yao2017deepsense, wang2019deep}. However, most existing solutions rely on centrally collected data, e.g., signal samples, including gyroscope and accelerometer time-series, collected from mobile users. 
Such labelled signal samples may contain private user information and may cause privacy concerns, according to data regulatory constraints that arise along with the widespread practice of data science, such as GDPR~\cite{albrecht2016gdpr}.
Due to this reason, to date few large-scale user activity dataset have been collected and made public, which hinders the development of HAR techniques.

Thanks to the rapid advancement of computational and storage capability on mobile devices, \emph{Federated learning} (FL) has emerged as an alternative distributed learning framework, which aims to train machine learning models based on decentralized data scattered on mobile devices without collecting them. 
In federated learning, a global model is downloaded by each mobile device and updated with its local data, while the local updates are aggregated into a renewed global model iteratively. 
Similar to other mobile applications such as keyboard input prediction~\cite{hard2018federated}, HAR is another well-motivated scenario that can benefit from federated learning~\cite{sozinov2018human}, 
which simplifies privacy management and gives each user a greater flexibility on controlling which local data samples are selected and how they should contribute to the overall application improvement.
Although Federated Learning has claimed decent performance in multiple tasks in the literature, e.g., image classification and keyboard input prediction~\cite{konevcny2016federated, hard2018federated}, it is still a question whether it can be used to solve HAR. Federated HAR has been tested on a simple deep neural network model with an accuracy reduction of up to 6\% reported by~\cite{sozinov2018human}, which is a significant degradation from centralized learning.
We point out that there are three major obstacles to performing HAR with federate learning:

\textit{First}, while federated learning is known to approximate centralized learning well if the training samples are independent and identically distributed (IID) across devices, such an IID assumption does not hold to activity signals. In fact, each user does not necessarily feature the same activity types---most users only perform a subset of all activities, e.g., one user may only have \{\textit{walking}, \textit{driving}\} recorded on his/her device, while another without any vehicle \blue{may only have \{\textit{walking}, \textit{biking}\}.} 
Therefore, the local training datasets have unbalanced label distributions. Such a heterogeneity in label distribution across users can cause serious performance degradation to federated learning~\cite{zhao2018federated}, and will exacerbate as the number of participating users increases.

\textit{Second}, even for the same activity, two users may exhibit dramatically different signal distributions.  The reason is because users may perform the same action in different styles. For example, two persons might have completely different walking patterns, one with large stride lengths, while the other with a high frequency and yet smaller step sizes. In other words, there exists a large degree of heterogeneity in the signal distributions of the same activity across users. We show through experiments that such heterogeneity in input signal distributions will also seriously affect the model accuracy achieved by federated learning.

\textit{Third}, a single global model found by federated learning can hardly adapt and generalize to individual users, especially to a new user with his or her own activity characteristics. 
To handle real-world HAR tasks, a personalized model combining the insights jointly mined from all users with a predictor specifically fit to its local data is desired for each user. Such personalization is especially desirable for a new user or an existing user with newly introduced activity types.

To solve these challenges, in this paper we propose Meta-HAR, a \emph{federated representation learning} framework for human activity recognition, where a shared, global deep embedding network is meta-trained by federated learning across users, while the signal representations given by the embedding network are fed into a separate classification network tailored to each device for personalized activity prediction. Such a representation learning framework is inspired by the fact that a good signal representation cannot be trained locally based on sparse data, but must take advantage of the abundant yet heterogeneous data residing on different devices, whereas the final personalized prediction at each user should augment the shared representation locally.

To train the shared embedding network, we treat each device (user) as a separate task, which possibly has its unique label distribution and input signal distributions. Inspired by model-agnostic meta-learning \red{\cite{finn2017model}}, we meta-train an embedding network that adapts to the distribution of the tasks instead of to each individual task, thus preserving the generalization ability to new tasks from this task distribution. Specifically, we adopt a federated version of \emph{Reptile}~\cite{jiang2019improving}, a first-order meta-learning algorithm~\cite{nichol2018first}, to train the embedding network. Each device iteratively updates the embedding network with its local dataset through a pairwise similarity loss and pushes the updated embedding network to the server for aggregation. 
By minimizing the pairwise loss instead of cross-entropy loss, samples from the same class are encouraged to cluster in the embedded space, while those from different classes are pushed apart.
Since every pair of samples yields a loss value for optimization, we have effectively bootstrapped the sparse data on each device into a larger amount of training samples.  
We show that the shared embedding network trained with this method is robust to the heterogeneous data distributions across users.
  
Once a generalizable signal representation is acquired, activity types are predicted on a device through a two-stage adaptation procedure.
We first fine-tune the embedding network on the local dataset of the device with pairwise loss. We then add a client-specific output layer on top of the embedding network for each user for activity classification, and fine-tune the embedding network and the output layer jointly on the same local data. 
\red{We show that the embedding network can be sufficiently fine-tuned on the small local dataset, and that the personalized models can even outperform a centrally trained model sometimes, especially for new users, due to the effective local adaptation.}

We have performed extensive evaluation of the proposed Meta-HAR on two publicly available datasets: 1) Heterogeneous Human Activity Recognition (HHAR) dataset~\cite{stisen2015smart} with 9 users and 6 activities, and 2) USC-HAD~\cite{zhang2012usc} dataset, with 14 users and 6 different activities, as well as on a much larger newly collected dataset\footnote{We open source the collected dataset
and all source code on Github: \url{https://github.com/Chain123/Meta-HAR}} with 48 users and 6 different activities. 
Note that the two publicly available datasets were created in carefully controlled environments such that each user has all activity types and a balanced label distribution. To mimic the real-world scenario, for these two datasets, we randomly removed several activities from each user to simulate the case of Non-IID label distributions. 
In each experiment, we randomly left several users out which served as the \emph{meta-test users} to test the generalizability of Meta-HAR to new users, and trained our model on the remaining \emph{meta-train users} using the proposed method. For each meta-train user, a portion of its local data was also left out for testing. We repeated every experiment 5 times and averaged the results.  
We have achieved test accuracies of \red{92.5\%}, \red{98.39\%} and \red{91.07\%}, \red{93.79\%} for meta-test users and meta-train users on HHAR and USC-HAD datasets, respectively. 
On the larger collected dataset, similar test accuracies of \red{93.29\%} and \red{90.76\%} are observed on meta-test users and meta-train users, demonstrating the scaling capability of Meta-HAR. Results on these datasets suggest that Meta-HAR clearly outperforms FedReptile~\cite{jiang2019improving}. 
We also merged the two publicly available datasets to stress-test Meta-HAR under a significantly heterogeneous and unbalanced scenario with 23 users and 7 different activities in total (with 5 overlapping activities in both datasets). 
In this case, we show that the proposed Meta-HAR significantly outperforms FedReptile. 

\section{Problem and Motivation}
\label{sec:problem}

Human activity recognition (HAR) aims to classify multitudes of sensor readings on mobile devices (e.g. signal segments from the gyroscope and accelerometer) into human activity types.
In this section, we introduce the HAR problem in a federated learning setting where sensor data cannot be centrally collected, together with the new challenges that HAR has posed to federated learning.

Suppose there are $n$ participating mobile devices (or users). 
Let the local dataset of user $i$ be represented by 
$D_i = \{(s_{ij}, a_{ij})|a_{ij} \in A^i, j = 1,2, \ldots, N_i\}$, where $s_{ij}$ is the \red{sensor signal of the} $j$-th sample in $D_i$, and $a_{ij}$ is its corresponding activity label. $N_i = |D_i|$ represents the number of samples in $D_i$, while $A^i$ denotes the set of activity types observed at user $i$. Notice that each user may have different activity types, i.e., $A^i$ is different across users. The ultimate goal is to solve the activity recognition problem on each individual user. 

A naive idea is to perform local supervised learning based on $D_i$ for each device separately. However, the pitfall here is that from the perspective of each user, the scheme has failed to leverage the vast amount of data residing on other users.  Moreover, since the activity types on each user are potentially sparse, a locally trained model can not make predictions about activities new to the user.

\subsection{Challenges to Federated HAR}
A seemingly plausible solution is federated learning, e.g., Federated Averaging (FedAvg)~\cite{mcmahan2016communication}, which is able to learn a global model on decentralized datasets residing on mobile users. However, this scheme does not work well for the HAR problem, mainly due to the heterogeneity that exists in both label and input signal distributions. 
First, it is shown that in image classification, the heterogeneity in label distribution among users, which is also referred to as the issue of Non-IID and unbalanced data, would cause substantial performance degradation to federated learning~\cite{zhao2018federated}. 
Aside from Non-IID label distribution, in HAR, users have heterogeneous input signal distributions even for the same activity, which can also cause performance degradation, but has not been reported in literature. 

Here, we demonstrate with experiments on HHAR dataset~\cite{stisen2015smart} that the heterogeneity in signal distributions alone can cause significant performance degradation. 
Note that in this dataset, each user has IID activity types (IID labels).
We split the data for each user into a train set (80\%) and a test set (20\%). The neural network model used for HAR task is shown in \red{Fig.~\ref{fig:structure}} and will be introduced in detail in Section \ref{sec:sys}. Three schemes are evaluated:
\begin{itemize} 
\item \textbf{Central}: Collecting all data on a server and train the HAR model centrally. 
\item \textbf{FedAvg-User}: \textit{FedAvg} is applied to learn a global HAR model across users. 
\item \textbf{FedAvg-Shuffle}: \textit{FedAvg} is applied to learn a global model, where all the samples are first collected and shuffled on a server then redistributed to all users.
\end{itemize}

\begin{figure} 
  \centering
  \includegraphics[width=2.6in]{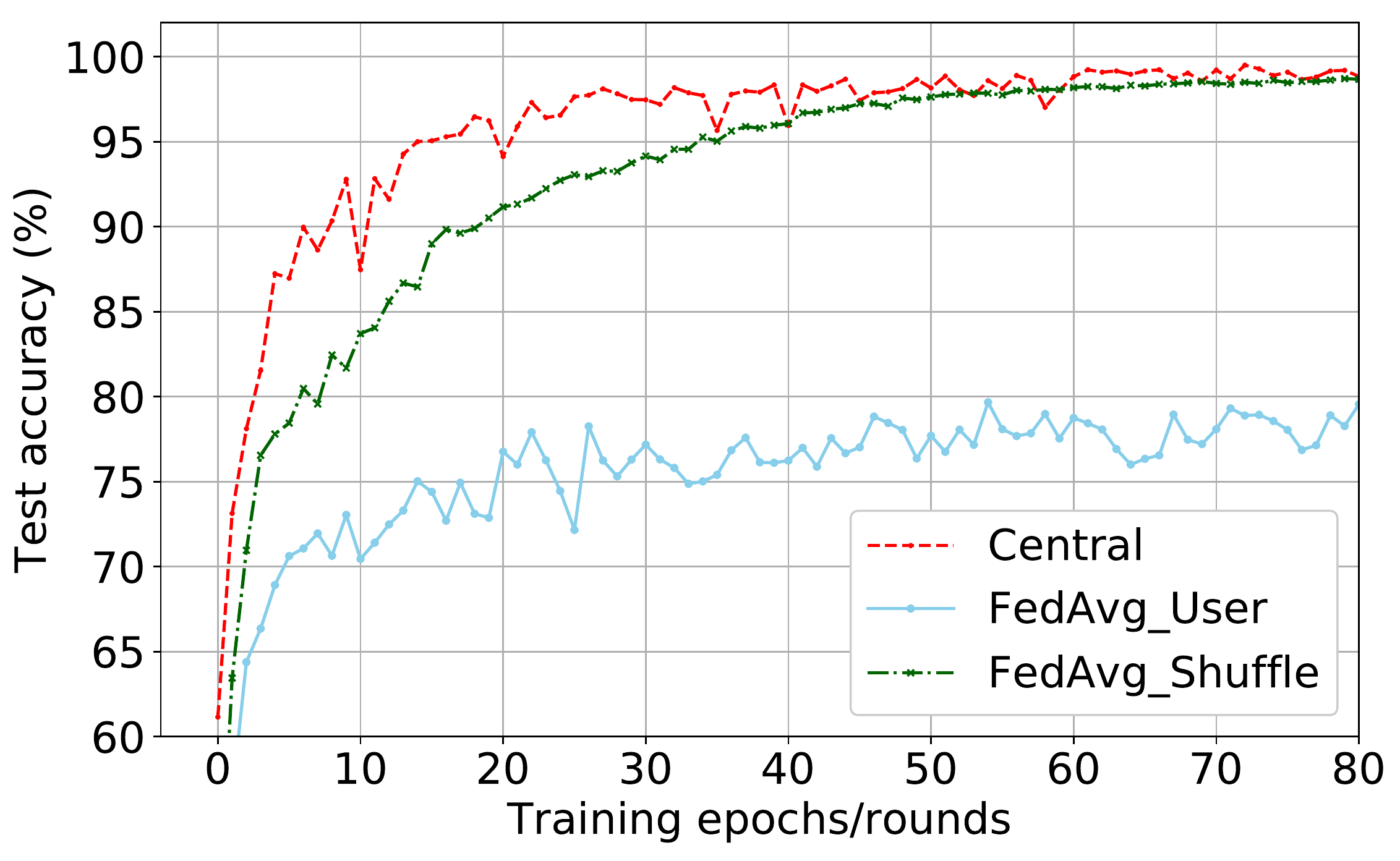}  
  \caption{Experimental results that demonstrate the insufficiency of \emph{FedAvg} on the HHAR dataset.}
  \Description{Training curves.}
  \label{fig:demo}
\end{figure}

Note that the original HHAR dataset is collected in a controlled situation thus it does not suffer from heterogeneity in label distribution (Non-IID or unbalance), thus the only difference between \textbf{FedAvg-User} and \textbf{FedAvg-Shuffle} is whether there is heterogeneity in signal distribution across users. The results are shown in Fig.~\ref{fig:demo}, where the x-axis represents training epochs for \textbf{Central} approach and the federated update rounds for two \textbf{FedAvg} schemes. One can easily observe an accuracy reduction of \textbf{FedAvg-User} compared to \textbf{Central} from over 97\% to below 80\%. While, without heterogeneity in signal \textbf{FedAvg-Shuffle} can learn a model that has performance close to centrally trained model. 
In our experiment, the performance reduction from Central model to FedAvg approach is nearly 20\%, much higher than that reported in previous work~\cite{sozinov2018human}, which is 6\%, due to the fact that~\cite{sozinov2018human} adopts generalizable handcrafted features and simple classification models, softmax regression and a DNN model. However, the performance of the state-of-the-art model for HAR under federated learning setting remains unknown and is worthy of further study. 

\begin{figure}[tbp]
  \centerline{
    \subfigure[``Stair-down'' activity data from all 9 users (a-i) in HHAR dataset.]{
      \includegraphics[width=1.5in, height=1.45in]{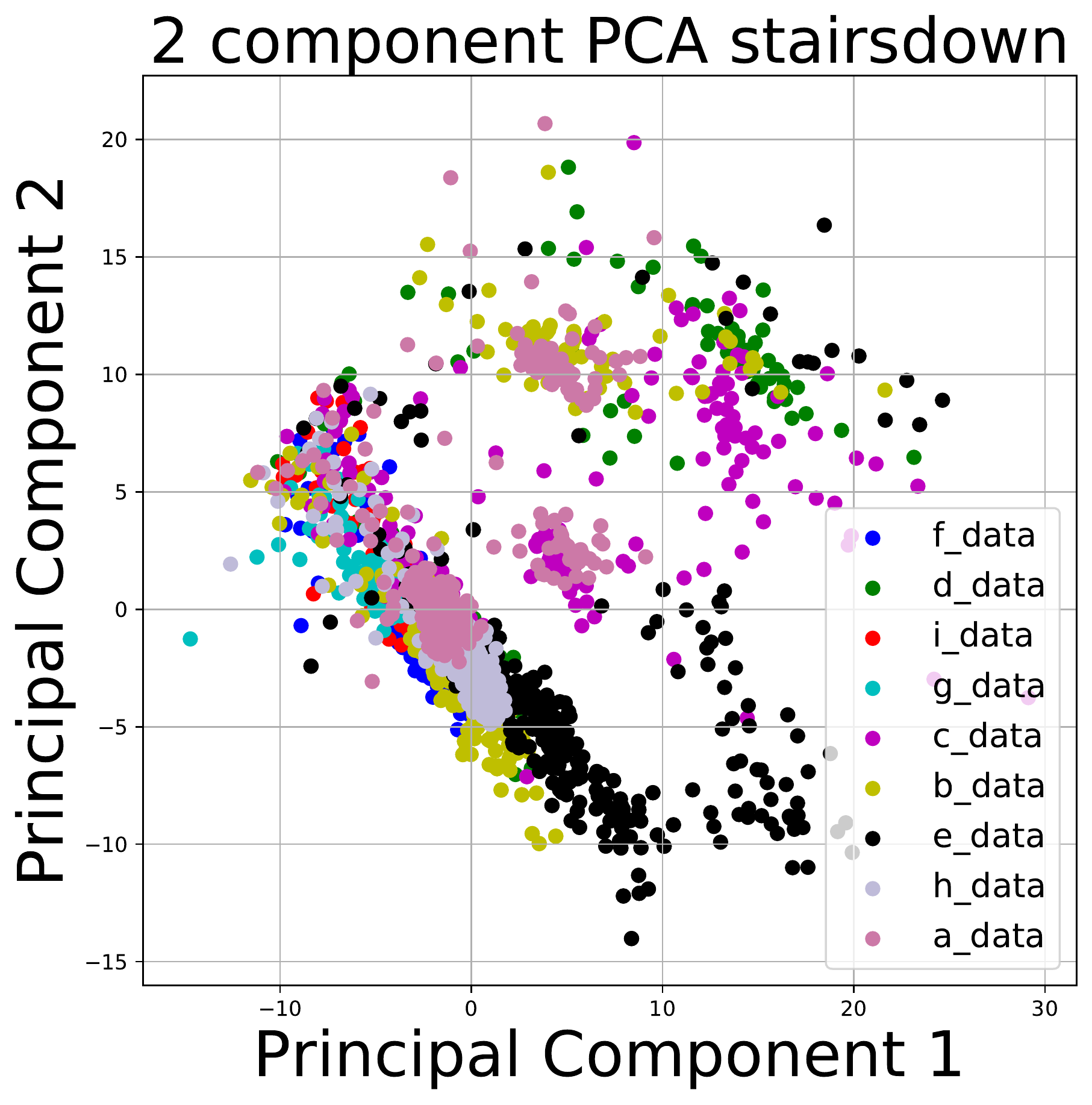}
      \label{fig:pca_act}
    }
    \subfigure[Different activities from user $h$.]{
      \includegraphics[width=1.5in, height=1.45in]{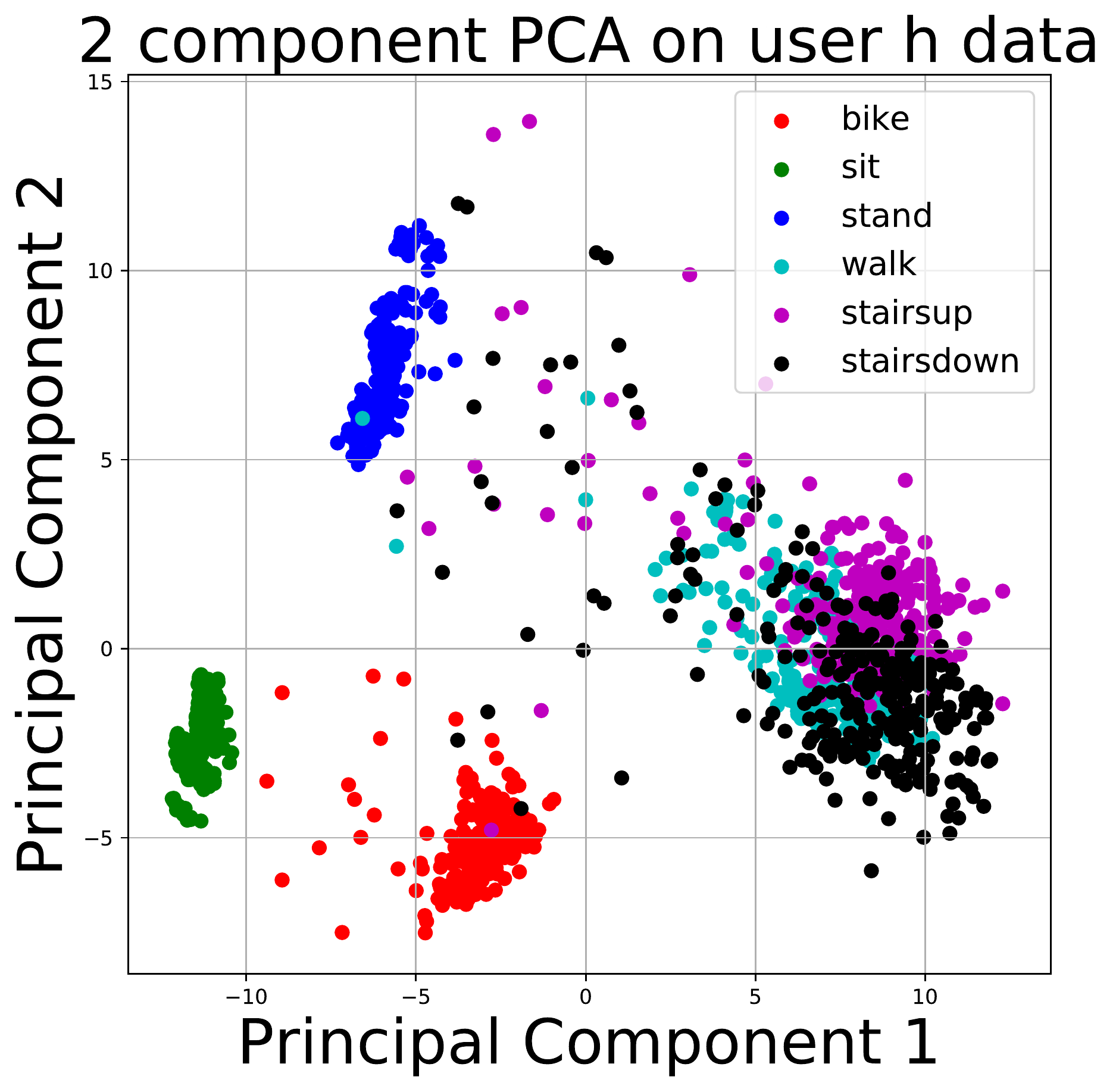}
      \label{fig:pca_user}
    }
  }
  \caption{(a) Distribution of samples from activity ``Stair-down'' for all the users in HHAR dataset; (b) Distributions of samples from all activities of user $h$.}
  \Description{Sample distributions.}
  \label{fig:pca}
\end{figure}

To further show the existence of heterogeneity among users even for the same activity, we extract traditional handcrafted features~\cite{sozinov2018human} (mean, standard deviation, maximum, minimum of signal amplitudes on each axis) and use PCA to visualize the sample distribution from HHAR dataset. The results are shown in Fig.~\ref{fig:pca}, as we can see in Fig.~\ref{fig:pca_user}, samples from different activities of user $h$ are well clustered which shows the efficiency of the handcrafted features. However, samples for the same activity ``Stair-down'' is also clustered by different users as shown in Fig.~\ref{fig:pca_act}, which demonstrates the heterogeneity in signal distribution across users. 

Aside from the performance issues we discussed above, there are also practical factors that shows the disadvantages of federated learning for our problem. 
First, without collecting user data, it can be hard to know the global activity set, $A = A^1 \cup A^2 \ldots A^n$, $n$ is the number of users, for the classification problem, which determines the output dimension of the global model. 
Furthermore, to train the model locally on user data, we need to unify the labels across different users, for example, we need to make sure label ``0'' represents the same activity across different users in federated HAR problem. This work is tedious and time-consuming. 
Finally, solving a classification problem with a tremendous number of classes is harder than a problem with fewer categories.

\blue{Therefore, a novel framework which is capable of overcoming the heterogeneity in both label and input signal distributions is desired for successful application of federated learning in human activity recognition.}


\section{Methods}
\label{sec:sys}
In this section we describe \textit{Meta-HAR}, our proposed federated representation learning framework for solve the HAR problem without centrally collecting the data. Instead of training a global classification network for all activity types, Meta-HAR first learns a common deep representation model (or a signal embedding network) parameterized by $\Theta$, through a model-agnostic meta-learning framework across all users. 
The goal of the embedding network is to embed any given input signal, regardless of its activity type, into a fixed length vector, which is fed into a classifier separately learned at each individual user to conform to its own activity set and output dimension. Such a design avoids the complexity to train a large global classifier for the global activity set and eliminates the need to unify labels across different users, as has been mentioned in Section~\ref{sec:problem}. 

In the following, we first provide an overview of the training procedure of the embedding network, followed by a description of the neural architectures of the embedding network adopted by Meta-HAR and its local training procedure. Finally, we present the procedures for model personalization and activity inference at the users. The overall workflow of Meta-HAR is shown in Algorithm~\ref{alg:fed_rep}.

\begin{figure}[t]
  \centering
  \includegraphics[width=3.3 in]{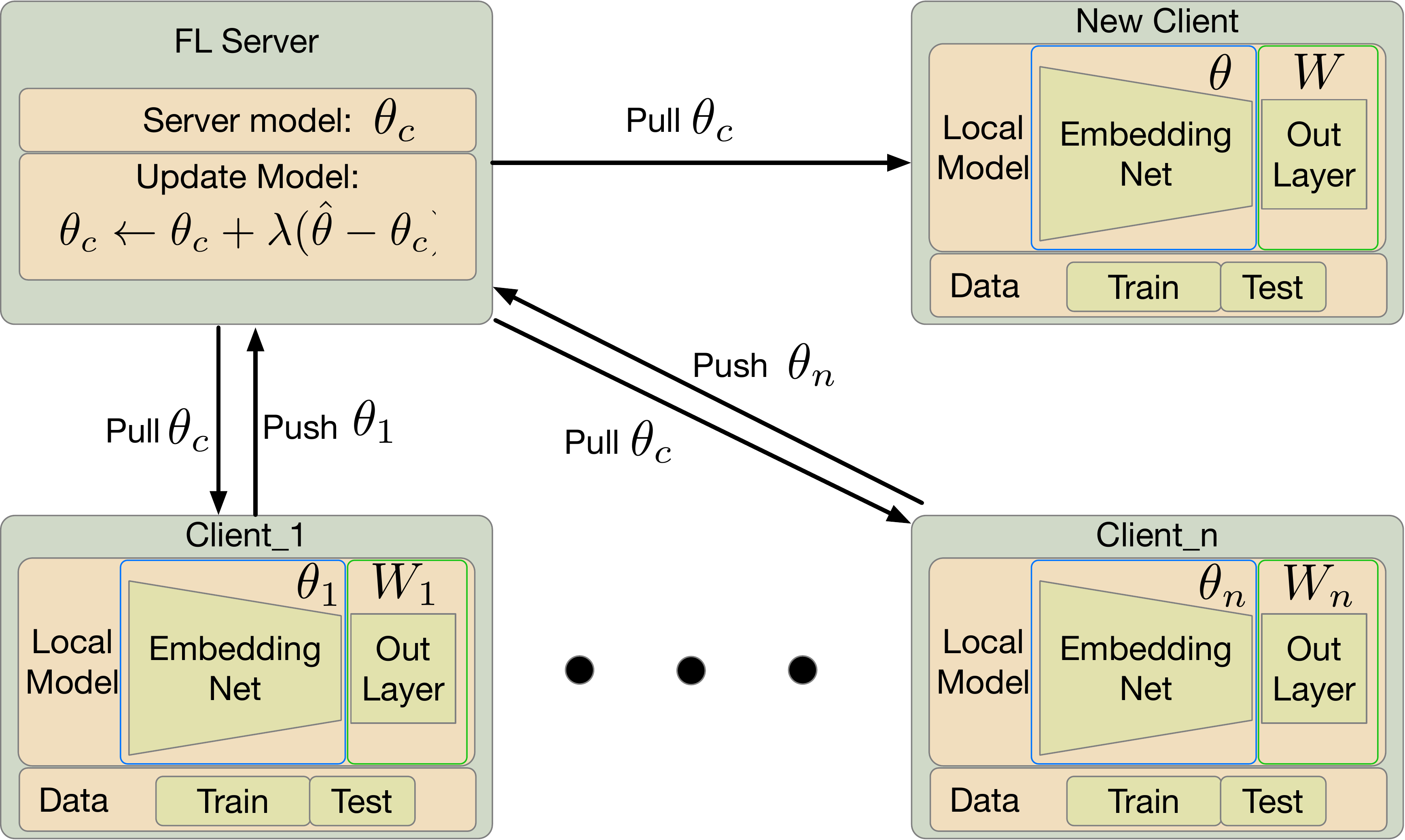} 
  \caption{An overview of the Meta-HAR framework. \red{In our system, there is a FL server, $n$ meta-train users and a new user as meta-test user. Each user holds a local dataset (which is further divided into train and test parts), a classifier parameterized by $\{\Theta, W\}$ where $\Theta$ represents the parameters of the embedding network and $W$ represents the output layer. Note that, only the embedding network $\Theta$ is meta-learned in a federated manner.}}
  \Description{System overview.}
  \label{fig:fed-meta}
  \vspace{-3mm}
\end{figure}

\subsection{Federated Representation Learning}
\label{sec:metalearn}
We first introduce our method to meta-learn the embedding network in a federated manner in order to achieve a strong generalization ability while leveraging heterogeneous data.
To introduce robustness in the presence of signal and label heterogeneity, the global embedding network is not trained by supervised classification tasks, but is meta-trained by pairwise comparison tasks, which compare whether two segments of signals belong to the same type of activities or not.

In federated HAR, although each user may have a local dataset with different activity types and signal distributions, yet the local task of activity classification is conceptually similar among different users. 
This is highly similar to the problem of optimization-based meta-learning~\cite{ravi2016optimization}, that attempts to learn an initialization of a model from multiple similar tasks, which can then be adapted to a new task at inference time through a fine-tuning stage, also known as \emph{adaptation}. Model-Agnostic Meta-Learning (MAML) is an emerging approach to learning to learn, whose goal is to train a model on a variety of similar learning tasks so that it can solve new learning tasks with only a small number of training samples and training steps. 
Therefore, we treat the activity recognition problem at each individual user as a separate task. Each task has its unique input signal distributions, a different output activity set and even a different number of activity types. Yet, each task is assumed to be sampled from $p(\mathcal T)$, a global distribution of tasks. According to optimization-based meta-learning, e.g., MAML~\cite{finn2017model, nichol2018first}, it is possible to train a model through a proper optimization procedure, such that the model adapts to the underlying distribution of tasks $p(\mathcal T)$ and thus can generalize to any new task sampled from $p(\mathcal T)$. 

Motivated by~\cite{jiang2019improving} which points out that FedAvg is a special case of Reptile, a scalable first-order meta-learning algorithm, we adopt a federated version of Reptile~\cite{nichol2018first}, which we call Federated Reptile (FedReptile), to meta-train the embedding network $\Theta$ in a federated manner among decentralized users.
The workflow of using FedReptile to train the embedding network, with parameters $\Theta$, is depicted in Fig.~\ref{fig:fed-meta}, in which there are $n$ users participating in training, a Federated Learning server (FL server) and a new user for testing of generalization ability. 
Each user holds a local dataset, pulls model parameters from the FL server, updates them using its local dataset, and then pushes those updates to the FL server. The FL server is responsible for coordinating the collaboration process of all users and updating the global model.  

As shown in Algorithm~\ref{alg:fed_rep}, suppose $n$ users are involved with local datasets $\{D_i\}, i = 1, 2, \ldots n$. In each round, a subset of users, $U$, is selected to update the global parameters. (According to~\cite{bonawitz2019towards}, in each round of Federated Learning, only a subset of users will participate to avoid excessive waiting time in a distributed setting.) Every user $i \in U$ first pulls the current parameters $\Theta_c$ from the FL server and performs $m$ epochs of local training before pushing the updated model $\Theta_i$ to the FL server. The FL server then averages the collected updates $\Theta_i$ to update the global model $\Theta_c$ as follows:. 
\begin{equation} \label{eq:rep_mean}
\hat{\Theta} = \frac{1}{|U|} \sum_{j \in U} \Theta_j,
\end{equation}
\begin{equation} \label{eq:rep_update}
\Theta_c = \Theta_c + \lambda(\hat{\Theta} - \Theta_c). 
\end{equation}

Finally, a personalization procedure is performed to obtain personalized models for all users, which we will discuss in detail in Section~\ref{sec:finetune}. 

\SetKwInput{KwInput}{Input}
\SetKwInput{KwOutput}{Output}

\SetKwInput{KwMetaTraining}{Meta-training}
\SetKwInput{KwPersonalization}{Personalization}

\begin{algorithm}[tb]
\DontPrintSemicolon
\caption{The Meta-HAR Algorithm}
\label{alg:fed_rep}
\KwInput{$n$ users with local train datasets $D = \{D_i\}, i= 1,\ldots,n$. A FL server with initialized embedding network $\Theta_c$.}
\KwOutput{Personalized HAR models for every user.}
\KwMetaTraining{}
\For{round = 1,2,3 ...}   
{ 
  Randomly select a subset $U$ of users. \\
  \For{User $j$ in $U$}
  {
    Pull model parameters $\Theta_c$ from server.  \\
    Train \red{$m (\geq 1)$} epochs of the embedding network on local dataset $D_j$ through pairwise loss, get locally updated parameters $\Theta_j$. \\
    Push the updated parameters $\Theta_j$ to server. \\
  }
  FL server update central model : $\Theta_c \gets \Theta_c + \lambda(\hat{\Theta} - \Theta_c)$ 
  where $\hat{\Theta} = \frac{1}{|U|} \sum_{j \in U} \Theta_j$
}
\KwPersonalization{}
\For{user $j$ in all users}
{
  Pull parameters of embedding network $\Theta_c$ from FL server.  \\
  Fine-tune $\Theta_c$ with pairwise loss on local dataset to obtain local embedding network $\Theta_j$.  \\
  Further fine-tune local classifier $\{\Theta_j, W_j\}$ with cross-entropy loss on local dataset. \\
  Return personalized classification model $\{\Theta_j^t, W_j^t\}$ for user $j$.  \\
}
\end{algorithm}

\subsection{Embedding Network Architecture and Local Training}
\label{sec:neuralnet}
In this subsection, we briefly introduce the structure of the embedding network and how the it is locally updated by individual users. 

Similar to prior deep neural networks for HAR (e.g.,\red{\cite{yao2017deepsense,nweke2018deep,yao2017deepiot}}), our embedding network $\Theta$ involves the combined use of convolutional networks and recurrent networks. However, $\Theta$ is not locally updated by minimizing the local classification loss (cross-entropy loss). 
Instead, a weight-sharing siamese network is used to predict whether two signal samples are of the same class or not. We now describe the embedding network architecture and local training methods. 

{\bf Embedding Network Architecture.} Similar to previous deep learning methods, we preprocess the raw readings from sensors before feeding them into a neural network. We first compute the amplitude series to serve as an additional input dimension for each sensor. For example, readings from motion sensor \textbf{Gyroscope} $s_g$ has three dimensions $s_g = \{ s_{gx}, s_{gy}, s_{gz}\}$. The additional amplitude axis is defined as $\sqrt{s_{gx}^2 + s_{gy}^2 + s_{gz}^2}$. 
\red{To model sequential dependencies of signal series, we split the sensor data into $k$ time intervals, each of width $\tau$.} 
After extending the amplitude axis \red{and segmenting the sequential signals}, we apply a Fourier transformation to each axis \red{of each segmented data block}, extracting the frequency domain representation. \red{Finally, we stack all the outputs of Fourier transformation, magnitudes and their corresponding frequencies, into a tensor of shape $k \times 2(d_g + 1) \times f$, where $d_g$ is the dimensionality of the sensor $s_g$, which is 3 in our example, and $f$ is the length of frequency domain representations.} 
The result is then fed to the embedding network $\Theta$.   

\begin{figure}. 
  \centering
  \includegraphics[width=2.6in]{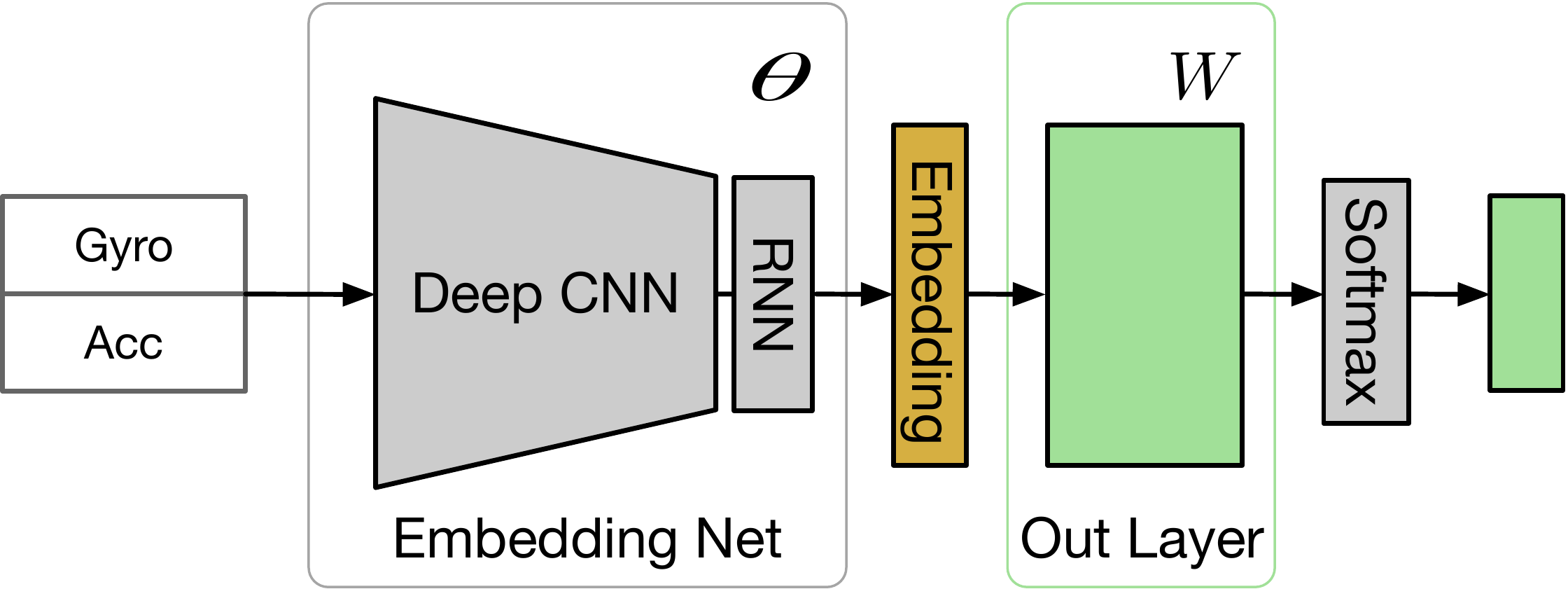}
  \caption{Deep classifier structure for HAR. The classifier consists of an embedding net and a fully connected output layer parameterized by $\Theta$ and $W$, respectively.}
  \label{fig:structure}
  \Description{HAR model architecture.}
\end{figure}

Similar to DeepSense \cite{yao2017deepsense}, the embedding network leverages both CNN and RNN to process sensor readings. As shown in Fig.~\ref{fig:structure}, multiple convolutional layers are applied to the processed sensor signals (from the Gyroscope and Accelerometer) to model spatial relevance among different axises of the same sensor as well as relevance across sensors. \red{To be specific, each of the $k$ data blocks of a given input sample, \red{we first} apply convolution to the stacked pair of magnitudes and frequencies along each axis of each sensor signal. Then, another convolutional layer is applied to all axes within the same sensor. Finally, we fuse the convolution outputs of different sensors through the last convolution layer. } 
\red{Then two Long Short-Term Memory (LSTM) layers are used to extract temporal relevance of the $k$ CNN outputs and output a fixed length embedding vector.}


Thereafter, shown in Fig.~\ref{fig:structure}, by adding a fully connected output layer on top of the embedding network, we can build a classifier to get the predicted probabilities for each category with a \textit{Softmax} function.

{\bf Training with Pairwise Loss.}
\red{
There are two loss functions that can be used to train the embedding network locally on the user side: \emph{cross-entropy} loss and \emph{pairwise} loss. 
Categorical cross-entropy loss, defined as: 
\begin{equation} \label{eq:xe}
H(p, q) = - \sum_{j=1}^{M}p_jlog(q_j),
\end{equation}
where $M$ represents the dimensionality of the two discrete probability distributions, measures the distance between a proposed probability distribution $q$ and target distribution $p$, is often used for multi-class classification problems. In our setting, the local HAR task on each user is a multi-class classification problem with the output dimension of $|A^i|$ for user $i$. However, as mentioned in Section~\ref{sec:problem}, different users have different local activity set. To leverage cross-entropy loss under the federated learning setting, one need to figure out the global activity set and unify the labels across different users so that local models have the same output layer $W$, which is time-consuming and hard to scale to new activity types. 
}
  
\red{
  Pairwise loss, on the other hand, does not require local users to be aware of the global activity set, thus is flexible and scalable to new activity types. Furthermore, pairwise loss encourages the clustering of sample embeddings in a real space, bootstraps the training set and makes the embedding network more robust to heterogeneous inputs. Well clustered embeddings can make the subsequent classification task much easier. Specifically, for a given pair of input samples $\{(e_i, a_i), (e_j, a_j) \}$ where $e_i, e_j$ are the embeddings output from embedding network for sample $i, j$ and $a_i, a_j$ are their corresponding labels. First, we use the cosine distance, defined as } 
\begin{equation} \label{eq:cosine}
\varphi_{ij} = \frac{e_i^T e_j}{|e_i| \cdot |e_j|}
\end{equation}
\red{to measures the similarity between any two embedding vectors.} 
Then, we the pairwise loss for sample pair $(i, j)$ is defined as 
\begin{equation} \label{eq:loss_p}
l_{i,j} = - \delta(a_i, a_j)log(\sigma( \varphi_{ij})) - (1 - \delta(a_i, a_j))log(1 - \sigma(\varphi_{ij}))
\end{equation}
where $\sigma(x) = 1/(1+ e^{-kx})$ is the logistic sigmoid function with a tunable parameter $k$, $\delta(a_i, a_j) = 1$ if $a_i=a_j$, otherwise $\delta(a_i, a_j) = 0$. Obviously, by minimizing the $l_{i,j}$, the cosine similarity $\varphi_{ij}$ between the embedded vectors $e_i$ and $e_j$ will reach the maximum if they are from the same class, i.e., $a_i = a_j$, and $\varphi_{ij}$ will reach the minimum if they are from different classes, i.e., $a_i \ne a_j$. 

\red{
To fit the model on a given dataset, we can sample a batch of $B$ pairwise samples, each in the form of $(s_i, s_j, a_i, a_j)$ to perform batched learning. For each training sample $(s_i, s_j, a_i, a_j)$, a weight-sharing siamese network is used to get the embeddings of the two input signals at the same time, that is, $e_i=\Theta(s_i)$ and $e_j=\Theta(s_j)$. Then the parameters of embedding network $\Theta$ will be updated by error back-propagation according to~\eqref{eq:loss_p}. 
}

\subsection{Personalized Classification}
\label{sec:finetune}

Finally, to solve the activity recognition problem on each user $i$, we introduce a personalized output layer parameterized by $W_i$, which together with the global embedding network $\Theta$ forms a classification model that conforms to the local output dimension and activity set at user $i$. However, the meta-learned global embedding network $\Theta_c$, when transferred to user $i$, only serves as an initialization of the embedding network to be used by user $i$.

For each user $i$, we use a two-step fine-tuning strategy to adapt the global embedding network $\Theta_c$ into its personalized local classifier. 
First, the embedding network $\Theta_c$ pulled from the server is fine-tuned on the local training set of user $i$ to obtain $\Theta_i$. The fine-tuning of the embedding network is also performed with the pairwise loss given by Equation~\ref{eq:loss_p}. 
The embedding network fine-tuned this way is able to encourage clustering of embedded signal representations based on the local dataset with user-specific activity types.     

The output of the local embedding network is then fed into a fully connected layer for activity classification. The weight parameters of the output layer for user $i$ are represented by $W_i$ with shape $(|E| \times |A_i|)$ where $|E|$ is the size of the output embedding vector, i.e., the dimension of the embedded space, while $A_i$ is the activity set of user $i$. Then, the output layer parameters $W_i$ together with the embedding network are fine-tuned with the cross-entropy loss. That is, in the second stage of fine-tuning, back-propagation is applied to $\{ \Theta_i, W_i\}$ to minimize the classification error. 

By leveraging a user-specific output layer, we have simplified the local classification by reducing its expected number of output categories in contrast to the global activity recognition problem for all users. 
On the other hand, as the embedding network is collaboratively learned across all users on abundant data and on all activity types, local classifiers built on top of the embedding network are capable of dealing with activities that have seldom seen by the user before. 
In other words, a well learned signal representation network has reduced the demand on the number of local samples required to fine-tune the local classifier.


\section{Experiments}
\label{sec:exp}
In this section, extensive experiments on three datasets, two public and one collected, are conducted to evaluate our proposed \textit{Meta-HAR} framework. 
Comparisons with several alternative baselines are also included to demonstrate the effectiveness of our method. 

\subsection{Datasets}
In this paper, 
two widely used public datasets: the Heterogeneous Human activity recognition (HHAR) dataset~\cite{stisen2015smart} with 9 users and 6 activities: \{\textit{Standing}, \textit{Sitting}, \textit{Walking}, \textit{Upstairs}, \textit{Downstairs} and \textit{Biking}\}, and the USC-HAD~\cite{zhang2012usc} dataset, with 14 users and 6 different activities \{\textit{Standing}, \textit{Sitting}, \textit{Walking}, \textit{Upstairs}, \textit{Downstairs} and \textit{Running} \} are adopted.   
Preprocessing is applied to maintain input consistency between the two public datasets and reformat the original sensor readings to fit our proposed architecture. First, we down-sample sensor readings in the HHAR dataset to a frequency of 25Hz. Signals in USC-HAD are down-sampled to 50Hz. Second, the long consecutive sensor data (up to 5 minutes in HHAR dataset) are segmented into several short samples, signals in HHAR dataset are segmented into $6$-second samples, while every $3$-second sensor readings in the USC-HAD dataset form a sample. This way, samples from different datasets would have the same input length.

Originally, users in the two public datasets have balanced samples for all activities. To mimic the real world, for each user we randomly remove 0 to 2 activities from its local dataset to simulate the scenario where datasets follow Non-IID distributions across users. Thereafter, the dataset has heterogeneity in both label and signal distributions. 
Finally, we merged these two public datasets to form an even more heterogeneous dataset with 23 users and 7 activities in total to stress-test the proposed method in comparison with baselines. 

\red{Aside from the two public datasets, we collected a new, larger HAR dataset involving 48 users and 6 types of activities, including \{\textit{Walking}, \textit{Biking}, (walking) \textit{Upstairs}, (walking) \textit{Downstairs}, \textit{Running} and \textit{Taking Bus/Taxi} \}. Our dataset was collected through an Android app specifically developed for activity signal logging. Participants can use any Android smartphone to choose an activity type, after which their signal samples are logged by the app while performing the activity.  Furthermore, there are no constraints or control on how users perform a certain activity. Therefore, our dataset is inherently more noisy and has more heterogeneity in terms of activity types, hardware devices, the position of the phone on the body, signal distributions, and sampling frequencies, etc., than prior datasets collected in controlled environments. Details on this collected dataset of 48 users are given in the Appendix \ref{sec:append}. }
\begin{table*}[tbp]
\caption{Test Results on HHAR, USC-HAD and collected Datasets. The number in the parenthesis denotes the fine-tuning epochs performed, e.g. Meta-HAR($x$) means Meta-HAR with $x$ epochs of fine-tuning on local datasets. All numbers are in percentage (\%).} 
\begin{center}
\begin{tabular}{c|cc|cc|cc}
\toprule
\multirow{2}{*}{Algorithms} & \multicolumn{2}{c}{HHAR Dataset} & \multicolumn{2}{c}{USC-HAD Dataset} & \multicolumn{2}{c}{Collected Dataset}\\ 
\cline{2-7} 
 & Meta-train user & Meta-test user & Meta-train user & Meta-test user & Meta-train user & Meta-test user\\
\midrule 
Central     & $98.55\pm0.11$ & $83.14\pm8.40$  & $99.31\pm0.14$ & $81.63\pm10.51$ & $90.18\pm0.14$ & $79.84\pm0.41$ \\
\midrule 
FedAvg      & $79.56\pm0.62$ & $66.79\pm1.84$  & $84.44\pm0.26$ & $80.24\pm1.54$ & $69.19 \pm5.02$ & $64.29\pm6.86$ \\
\midrule
FedReptile(1) & $87.16\pm0.29$ & $86.00\pm1.66$  & $87.96\pm0.21$ & $85.33\pm2.20$ & $88.29\pm0.66$ & $83.96\pm2.31$\\ 
FedReptile(2) & $92.64\pm0.26$ & $88.04\pm2.65$  & $91.02\pm0.30$ & $86.44\pm3.05$ & $90.84\pm0.18$ & $89.59\pm1.29$\\
FedReptile(3) & $95.70\pm0.25$ & $91.84\pm1.85$  & $\mathbf{93.98\pm0.31}$ & $89.17\pm2.26$ & $\mathbf{91.49\pm0.31}$ & $92.30\pm1.52$\\
\midrule
Meta-HAR(1)   & $98.32\pm0.06$ & $85.23\pm1.80$  & $92.01\pm0.16$ & $83.59\pm2.28$ & $89.16\pm0.76$ & $92.38\pm0.43$\\
Meta-HAR(2)   & $98.36\pm0.04$ & $91.25\pm1.82$  & $92.54\pm0.11$ & $90.19\pm2.81$ & $90.24\pm0.63$ & $\mathbf{93.33\pm0.80}$\\
Meta-HAR(3)   & $\mathbf{98.39\pm0.02}$ & $\mathbf{92.50\pm1.26}$  & $93.79\pm0.14$ & $\mathbf{91.07\pm1.74}$ & $90.76\pm0.67$ & $93.29\pm1.03$\\
\bottomrule
\end{tabular}
\label{table:Fed_heter}
\end{center}
\end{table*}


\subsection{Experiment Setups}
To begin with, we split all the users in a dataset into \textbf{Meta-train users}, which participate in the meta-learning process, and \textbf{Meta-test users}, which served as new users for testing the generalization ability of the meta-learned model. 
To be specific, on the HHAR or USC-HAR dataset, we randomly select \emph{one user} as the meta-test user. 
On the merged dataset, \emph{two users} are randomly selected as meta-test users from HHAR and USC-HAD datasets, respectively. For the collected dataset, \emph{five users} are selected as meta-test users.    
Our ultimate goal is to learn a personalized model for all users, meta-train and meta-test users, that can solve the local human activity recognition problem. 
To train the model and test the performance, we further split the local dataset of each user into a train set (80\%) and a test set (20\%). 
Specifically, the following schemes are evaluated to demonstrate the effectiveness of our proposed method: 
\begin{itemize} 
  \item \textbf{Central}: The HAR classification model is trained on all data samples collected on a server.  
  \item \textbf{FedAvg}: The original Federated Averaging method ~\cite{mcmahan2016communication}, where a global, shared classification model is learned in a federated manner.
  \item \textbf{FedReptile}: The Federated Reptile \cite{jiang2019improving} is applied to first meta-learn a global initialization of a classification model. Then personalization is achieved by fine-tuning the classification model with cross-entropy loss on each user.  
  \item \textbf{Meta-HAR}: The proposed \textit{Meta-HAR} framework. 
  \item \textbf{Meta-HAR-CE}: A variant of the proposed \textit{Meta-HAR}, where the embedding network is trained with a cross-entropy loss instead of pairwise loss.    
\end{itemize} 
Note that compared to \textbf{FedAvg}, \textbf{FedReptile} simply adds extra adaptation steps to generate personalized models. However, different from previous works, only the embedding network is federated learned in our proposed method.

Furthermore, the following fine-tuning strategies for personalization are evaluated and compared:
\begin{itemize} 
  \item \textbf{Separated}: The parameters of the embedding network and the output layer are fine-tuned independently. Specifically, the embedding network is first fine-tuned with pairwise loss, then we fix the embedding network to further fine-tune the output layer with cross-entropy loss. 
  \item \textbf{Merged}: The embedding network and the user-specific output layer are jointly fine-tuned with cross-entropy loss. 
  \item \textbf{Two-stage}: The proposed two-stage fine-tuning strategy. 
\end{itemize}

\red{Note that the FedAvg \cite{mcmahan2016communication} scheme is well designed to overcome systematic challenges such as communication efficiency and limited computational power on mobile devices, etc. 
Our proposed framework follows the same system design as FedAvg, i.e, we have similar system efficiency and communication cost as FedAvg does. 
In this work, our goal is to solve the data heterogeneity problem posed by federated HAR, therefore, instead of addressing the communication challenges, we focus on the model performance for the HAR problem.}  
The performance of a model is evaluated with the averaged prediction accuracy which is defined as follows:
\begin{equation} \label{eq:Avg_acc}
Acc = \frac{1}{\sum_{i= 1}^{n} m_i}\sum_{i = 1}^{n} m_i \cdot acc_i,
\end{equation}
where $acc_i$ is the test accuracy on the $i$th user and $m_i$ is the number of test samples on that user. The averaged accuracies on both meta-train and meta-test users are used to evaluate the generalization ability to future data of users who participated in the meta-learning as well as to new users who have not participated in the embedding network training. 

In each experiment, we randomly select the meta-test and meta-train users to evaluate the performance for the proposed method and all baselines. We repeat the whole process 5 times on each dataset. The mean accuracies and their standard deviations are presented. 

\subsection{\blue{Implementation Details}}
Our models are implemented using Pytorch with python 3.6.  
All experiments are carried out on Tesla P40 GPUs with memory size of 22.38 GiB and 1.53 GHz memoryClock-Rate. ADAM optimizer\cite{kingma2014adam} with $\beta_1 = 0.9$, $\beta_2 = 0.98$ and $\epsilon = 1 \time 10^{-8}$ is used to update all network parameters. 
We use $k=10$ for the sigmoid function $\sigma(x) = 1/(1+ e^{-kx})$ to calculate pairwise loss defined in Equation~\eqref{eq:loss_p}. In federated learning procedure, we set $\lambda = 1.0$ and perform $m=2$ epochs of local training at each update round. 

Our human activity recognition model consists of multiple 1-D and 2-D convolutional layers all with 64 filters and two layers of LSTM layers. The size of the latent vector is set to be 100. We adopt dropout to prevent over-fitting in the training stage and the dropout rate is set to be 0.3. As the number of samples residing on the mobile devices of users is usually small, we adopt a batch with size 64 for local training.     

\subsection{Experimental results}
\textbf{Comparison between \textit{Central}, \textit{FedAvg}, \textit{Meta-HAR} and \textit{FedReptile}}.  
\blue{Table~\ref{table:Fed_heter} compares our proposed method with baseline approaches on the three datasets: two public datasets and one collected dataset. The results on the merged dataset are given in table~\ref{table:Fed_merge}. 
Comparing the \textit{Central} model with all other federated learning-based methods, we can see there is a great performance degradation from \textit{Central} to \textit{FedAvg} in terms of the activity prediction accuracy on all datasets. 
On the other hand, \textit{Meta-HAR} and \textit{FedReptile} can achieve comparable performance on meta-train users and are even able to outperform the \textit{Central} model by a great margin on meta-test users. This demonstrates the superior of personalized models over a single federated learned global model. 
The advantage of the personalized models, generated by \textit{Meta-HAR} and \textit{FedReptile}, over the single global model, learned through \textit{FedAvg}, can be contributed to the adaptation of global model to the local datasets.  
After fine-tuning, personalized models on the user side can be better fitted to the distributions of the corresponding local datasets. 
}

\blue{
Notice that on the USC-HAD dataset, there is a $5.33\%$ performance reduction from the \texttt{Central} model to the best-personalized models on meta-train users. This is because, on the USC-HAD dataset, each user only holds a small local dataset with only a few samples, therefore, it is hard to get a personalized model with a performance comparable to the \textit{Central} model. The insufficiency of local datasets on the mobile devices of users further motivates the need for federated learning where we can leverage all the data samples scattered on mobile devices. 
}

\blue{
Comparing the \textit{Meta-HAR} model with the \textit{FedReptile} approach, we can see these two methods achieved decent and close performance after fine-tuning on local datasets. \textit{FedReptile} can sometimes even be a slightly better than \textit{Meta-HAR} on meta-train users, e.g. on USC-HAD dataset.  
This is reasonable due to two factors: first, there are only a few heterogeneities in label and signal distributions on the simple public datasets, particularly on the USC-HAD dataset. 
Second, in meta-train users, all parameters including the last output layer of the local classification models are federated learned, i.e. they have better initial weights when performing personalization. 
However, on complex datasets with more heterogeneities and meta-test users, \textit{Meta-HAR} can significantly outperform \textit{FedReptile}, 
e.g. on the merged dataset, the \textit{Meta-HAR} model achieves an accuracy of $95.35\%$ greatly outperform \textit{FedReptile} which gives an accuracy of $70.65\%$ on meta-train users.    
On meta-test users, \textit{Meta-HAR} outperforms \textit{FedReptile} on all datasets, for example, on the merged dataset, \textit{Meta-HAR} achieves an averaged accuracy of $75.83\%$ and $90.23\%$ on meta-test users from HHAR and USC-HAD datasets, respectively, greatly outperform the accuracy achieved by \textit{FedReptile}, which is $69.4\%$ and $74.83\%$. 
This demonstrates our model's capability of handling datasets with high heterogeneity and can be easily adapted to new (meta-test) users. 
The advantages of \textit{Meta-HAR} over \textit{FedReptile} can be contributed to two main reasons. First, in our framework, only the embedding network is federated learned with pairwise loss, which is more robust to heterogeneity in both labels and signals. Second, with pairwise loss, we can boost the local dataset and improve the learning efficiency, which is suitable for a small dataset residing on user's mobile devices.        
}

\begin{table}[tbp]
\caption{Test Results on Merged Dataset for FedAvg, FedReptile, Meta-HAR-CE and Meta-HAR methods. The number in the parenthesis denotes the fine-tuning epochs performed. All numbers are in percentage (\%).}
\begin{center}
\scalebox{0.95}[0.95]
{
\begin{tabular}{cccc}
\toprule
Algorithms      & Meta-train & Meta-test (H) & Meta-test (U)   \\
\midrule 
FedAvg          & $48.97\pm0.62$ & $39.99\pm1.56$ & $52.71\pm1.97$  \\
\midrule    
FedReptile(1)     & $58.12\pm0.55$ & $59.52\pm1.87$ & $66.16\pm2.52$  \\
FedReptile(2)   & $65.83\pm0.65$ & $66.35\pm3.67$ & $71.95\pm2.72$  \\
FedReptile(3)   & $70.65\pm0.53$ & $69.40\pm2.64$ & $74.83\pm2.19$  \\

\midrule 
Meta-HAR-CE(1)    & $94.05\pm0.41$ & $62.69\pm1.90$ & $61.37\pm3.62$  \\
Meta-HAR-CE(2)    & $97.01\pm0.36$ & $62.74\pm3.00$ & $71.12\pm2.52$  \\
Meta-HAR-CE(3)    & $\mathbf{97.70\pm0.24}$ & $67.19\pm3.47$ & $80.96\pm2.95$  \\

\midrule 

Meta-HAR(1)     & $91.42\pm0.49$ & $47.28\pm1.65$ & $73.98\pm2.68$  \\
Meta-HAR(2)     & $93.64\pm0.39$ & $54.38\pm2.47$ & $87.00\pm2.88$  \\
Meta-HAR(3)     & $95.35\pm0.28$ & $\mathbf{75.83\pm2.27}$ & $\mathbf{90.23\pm2.44}$  \\
 \bottomrule
\end{tabular}
}
\label{table:Fed_merge}
\end{center}
\end{table} 

\textbf{Comparison between \textit{Meta-HAR} and \textit{Meta-HAR-CE}}. Comparison between our proposed method, \textit{Meta-HAR}, and its variant, \textit{Meta-HAR-CE}, is performed as an ablation test to further demonstrate the advantage we get by adopting pairwise loss to meta-train the embedding network. 
Show in table~\ref{table:Fed_merge},  
one can see, for most of the time, \textit{Meta-HAR} outperforms \textit{Meta-HAR-CE} with a great margin on meta-test users. On meta-train users, it achieved a slightly worse accuracy than \textit{Meta-HAR-CE} did. This makes sense, cause in the meta-training procedure of \textit{Meta-HAR-CE}, all parameters of the local classifier on meta-train users are already well trained on the local dataset. 
The higher performance achieved by \textit{Meta-HAR} over \textit{Meta-HAR-CE} on the meta-test user demonstrates the superior generalization ability of \textit{Meta-HAR}.  
\red{Furthermore, mentioned in Section \ref{sec:neuralnet}, \textit{Meta-HAR} is scalable to new activity types in federated learning setting while \textit{Meta-HAR-CE} can not due to fixed output dimensionality of cross-entropy loss.} \blue{These advantages of \textit{Meta-HAR} over \textit{Meta-HAR-CE} can all be contributed to the adoption of pairwise loss.}

\textbf{Evaluation of different fine-tune strategies}. 
Finally, we evaluate several fine-tuning strategies, to show the effectiveness of our proposed two-stage adaptation approach. Evaluations are done on the merged dataset. Results of proposed two-stage method, \textit{Merged} and \textit{Separated} approaches are presented in Table~\ref{table:Fed_tune}. All the strategies can achieve decent test accuracies on meta-train users, however, our two-stage fine-tuning method significantly outperforms other baseline approaches on the meta-test users by a great margin. 
\textit{Merged} and \textit{Separated} methods fine-tune both the embedding network and output layer the same way as we do in the two-stage method. However, the improvement is insignificant and thus cannot be used for fast adaptation to new users. The \textit{Merged} method achieved better performance than \textit{Separated} did which shows the advantage by jointly fine-tuning the embedding network and output layer.

\begin{table}[tbp]
\caption{Test Results of Meta-HAR with different fine-tune methods on Merged Dataset. The number in the parenthesis denotes the fine-tuning epochs performed. All numbers are in percentage (\%).}
\begin{center}
\begin{tabular}{cccc}
\toprule
Tune methods    & Meta-train & Meta-test (H) & Meta-test (U)   \\
\midrule 
Merged(1)     & $89.03\pm0.83$ & $47.60\pm2.07$ & $60.76\pm3.62$  \\
Merged(2)     & $91.84\pm0.62$ & $48.47\pm2.36$ & $69.79\pm2.62$  \\
Merged(3)   & $\mathbf{95.38\pm0.47}$ & $50.38\pm1.59$ & $80.19\pm3.39$  \\
\midrule  
Separated(1)    & $88.75\pm0.65$ & $47.22\pm2.54$ & $51.98\pm1.79$  \\
Separated(2)  & $90.47\pm0.69$ & $48.69\pm2.85$ & $64.44\pm1.16$  \\
Separated(3)  & $91.91\pm0.71$ & $50.98\pm2.97$ & $73.35\pm1.72$  \\
\midrule  
Two-stage(3)  & $95.35\pm0.28$ & $\mathbf{75.83\pm2.27}$ & $\mathbf{90.23\pm2.44}$  \\
\bottomrule
\end{tabular}
\label{table:Fed_tune}
\end{center}
\end{table}


\section{Related Work}
\label{sec:related}
\textbf{HAR with deep learning model.} 
There are several recent studies leveraging deep neural network models to different mobile sensing or HAR applications.  
Deep Boltzmann Machine is adopted in DeepEar \cite{lane2015deepear} to improve the performance of audio sensing tasks in an environment with background noise. 
DeepX \cite{lane2016deepx} and RedEye \cite{likamwa2016redeye} reduce the energy consumption of deep neural networks, based on software and hardware, respectively. 
IDNet \cite{gadaleta2016idnet} uses CNNs for the biometric gait analysis. 
RBM \cite{bhattacharya2016smart} and MultiRBM \cite{radu2016towards} combine deep Boltzmann Machine and Multimodal DBMs to boost performance of human activity recognition task. 
Deepsense \cite{yao2017deepsense} applied RNN on top of CNN to acquire the sequential information of the input sensor data. 

\textbf{Federated optimization.} Federated learning \cite{konevcny2016federated} aims to train a high-quality centralized model with data defining the optimization problem being unevenly distributed over large number of nodes. This data decentralization can help reduce data transmission and protect user privacy. 
Most of the recent studies focus on addressing the communication efficiency challenge faced by federated learning. FederatedAveraging \cite{mcmahan2016communication}, which combines local stochastic gradient descent (SGD) on client nodes with model averaging on the server side, is able to reduce communication rounds between clients and server. 
\citet{bonawitz2017practical} allows a server to sum up large vectors from mobile devices in a secure manner through a communication-efficient, failure-robust protocol. 
Differential privacy \cite{mcmahan2017learning} achieves user-level privacy protection for the federated averaging algorithm. \citet{mcmahan2016communication,sattler2019robust} also show the robustness of federated learning algorithm when clients hold Non-IID data. \citet{zhao2018federated} proposed to improve FederatedAveraging on Non-IID data by sharing a small subset of client data on the server side. 
However, sharing local samples is infeasible for real world federated learning application where the number of clients is extremely large and the data on client's side may update frequently, for example the log data of APPs installed on smart phones. 
\citet{sozinov2018human} first applied federated to the problem of human activity recognition. However, it simply applied FedAvg to HAR without trying to solve the new challenges posed by federated HAR.    
In this paper, apart from Non-IID and unbalanced distribution in label, we pointed out that there is also heterogeneity in signal distribution can cause performance degradation in Federated Learning. We propose to leverage model personalization with MAML algorithm on a global embedding network to addressing the challenges face by federated HAR.  

\red{\citet{smith2017federated} propose a Federated Multi-Task learning framework MOCHA to solve the general multi-task learning problem, which is a plausible solution to our problem. However, they focus on solving high communication cost, stragglers, and fault tolerance for distributed multi-task learning and did not discuss the case of heterogeneous input signal distribution and can not adapt to new arriving users due to the fact that there is no global shared representation model in MTL setting.} 
\blue{FAVOR \cite{wang2020optimizing} uses reinforcement learning to achieve intelligent device
selection to improve federated learning performance and overcome Non-IID data distribution over participants. However, it can not provide personalized models for all participant users. }

\textbf{Meta-learning.} 
Meta-learning aims to solve the problem of learning to learn~\cite{naik1992meta,thrun2012learning}. Early works focus on the design of meta-trainers, i.e., a model that learns how to train another model such that better performance can be achieved on a given task~\cite{bengio1992optimization,schmidhuber1992learning}. 
\citet{andrychowicz2016learning} adopts deep neural networks to train a meta-learner and proposes an optimizer-optimizee setup, where each component is learned with an iterative gradient-descent procedure. 
\citet{li2016learning} follows a guided policy search strategy and automatically learns the optimization procedure for updating a model. 
Model-agnostic meta-learning (MAML)~\cite{finn2017model} is another popular approach that does not impose a constraint on the architecture of the learner. 
\citet{ravi2016optimization} proposes an LSTM meta-learner to learn an optimization procedure for few-shot image classification. 
\citet{li2017meta} develops an SGD-like meta-learning process and also experiment on few-shot regression and reinforcement learning problems.
Reptile~\cite{nichol2018first}, i.e., the approach adopted in this paper, simplifies the learning process of MAML by conducting first-order gradient updates on the meta-learner. \citet{jiang2019improving} interpreted federated learning as a MAML algorithm and implement a federated version of the first-order MAML algorithm, Reptile. However, \citet{jiang2019improving} focus on the parameter tuning to get a global model which is readily to personalize. 

\textbf{Learn personal models.} 
Several algorithms have been developed for training personalized model on decentralized peer-to-peer network \cite{vanhaesebrouck2017decentralized}. Users with local datasets collaborate with each other through peer-to-peer exchange in the network in order to learn personalized model. Personalized model is convenient for local users, due to the fact that the newly generated data on the mobile devices are often consumed locally on that device for many applications, for example, users' interaction log with APP. However, the peer-to-peer network is unrealistic for the federated learning setting where the number of clients if extremely large and the relationships between clients are complicated and even dynamic. \citet{jiang2019improving} combined federated learning with MAML algorithm and shown with numerical results the benefits by adopting personalization for Federated Learning. 
However, previous only made incremental contribution to the original Federated Learning framework by simply adding a personalization step. On the other hand, our framework is more flexible, personal models can even have different output dimension for different clients.


\section{Conclusions}
\label{sec:conclude}

In this paper, we study the federated human activity recognition problem, which aims to train accurate personalized activity classification models for mobile users or devices, without centrally collecting their sensor data. The Non-IID activity type distribution across users as well as the heterogeneity in signal distribution across different users have posed significant challenges to federated learning for HAR. 
We propose a federated representation learning framework for HAR, namely \textit{Meta-HAR}, to  to meta-learn an embedding network in a federated manner, leveraging the heterogeneous yet abundant datasets residing on distributed mobile devices. 
A personalized model with user-specific output layer can then be obtained for each user through an adaptation strategy on top of the meta-learned embedding net. 

Extensive experiments on one collected dataset, two publicly available datasets and their merged dataset are conducted. Our approach significantly outperformed all baselines methods with a great margin. For example, on the merged dataset we outperform the baseline FedReptile method by \red{24.7\%}, \red{6.43\%} and \red{15.4\%} on meta-train users, HHAR meta-test users and USC meta-test users, respectively. \blue{We also make our newly collected dataset publicly available to facilitate future development in sensor-based human activity recognition.}

\bibliographystyle{ACM-Reference-Format}
\bibliography{acmart}

\appendix
\section{Collected Dataset}
\label{sec:append}

\begin{figure}[ht]
  \centering
  \includegraphics[width=1.6 in]{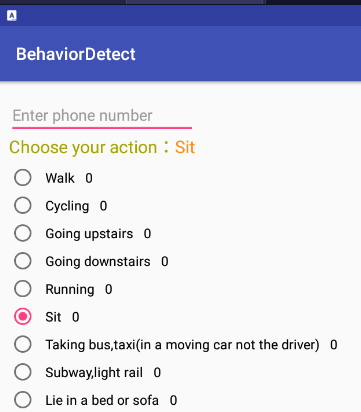}
  \caption{Screen shot of our developed Android app for sensor data logging. The number right after each action type represents the number of samples that have been collected and uploaded.}
  \Description{App screen shot.}
  \label{fig:app}
\end{figure}

\blue{We developed an Android app specifically designed for human activity sensor signal logging in a real-world scenario. The screenshot for the developed app is shown in Fig.~\ref{fig:app}, there are 9 types of activities that users can choose to perform, however, finally, we got enough data samples for only 6 of them: \{\textit{Walking}, \textit{Biking}, \textit{Upstairs}, \textit{Downstairs}, \textit{Running} and \textit{Taking Bus}\}. The published dataset is the same as we used in this paper. 
Every participant needs to go through the following steps to upload their sensor signals.}
\begin{itemize}
  \item Enter your phone numbers which served as your ID. 
  \item Choose an activity that you are going to perform and press the ``start'' bottom. 
  \item Performing corresponding activity until you are done. 
  \item Press ``Stop'' (the ``Start'' button becomes the ``Stop'' button after you pressed start.) The app will automatically upload collected sensor signals to a pre-assigned server. 
\end{itemize}
During data collection, every 7-second consecutive signals will be saved as one data sample. Normally, users wouldn't perform the corresponding activity immediately right after they pressed start. Therefore, we discard the signals of the first 5 seconds and the last 7 seconds to avoid bad samples and reduce sample noise. 

The sampling rate is set to be 25Hz, however, due to the heterogeneity among hardware devices and operating systems (customized Android systems and different versions of the Android system), we can not let the collected samples from different users be exactly 25 Hz. Therefore, we only kept samples with a sequence length between $[150, 200]$, which is a frequency between $[21.4Hz, 28.6Hz]$, and discard all samples that do not fall in this section. 
Finally, we select the first 150 data points of each sample so that we have the same input data shape across all users. 

Note that, unlike previous work, where the sensor signals of human activity are collected in a carefully designed and well-controlled environment \cite{zhang2012usc,stisen2015smart}. In our case, the users are totally free when they performing a certain activity. Therefore, our dataset is inherently noisier with more heterogeneities in terms of local activity types, hardware devices, local data unbalances, the position of the phone on the body, signal distribution, and sampling rate. To make sure, users are performing the correct chosen activity, we require users to upload a short video recording their surroundings and action to avoid cheating. 

\begin{figure}[h]
  \centerline{
    \subfigure[Histogram for the number of activity types of local datasets]{
      \includegraphics[width=1.45in]{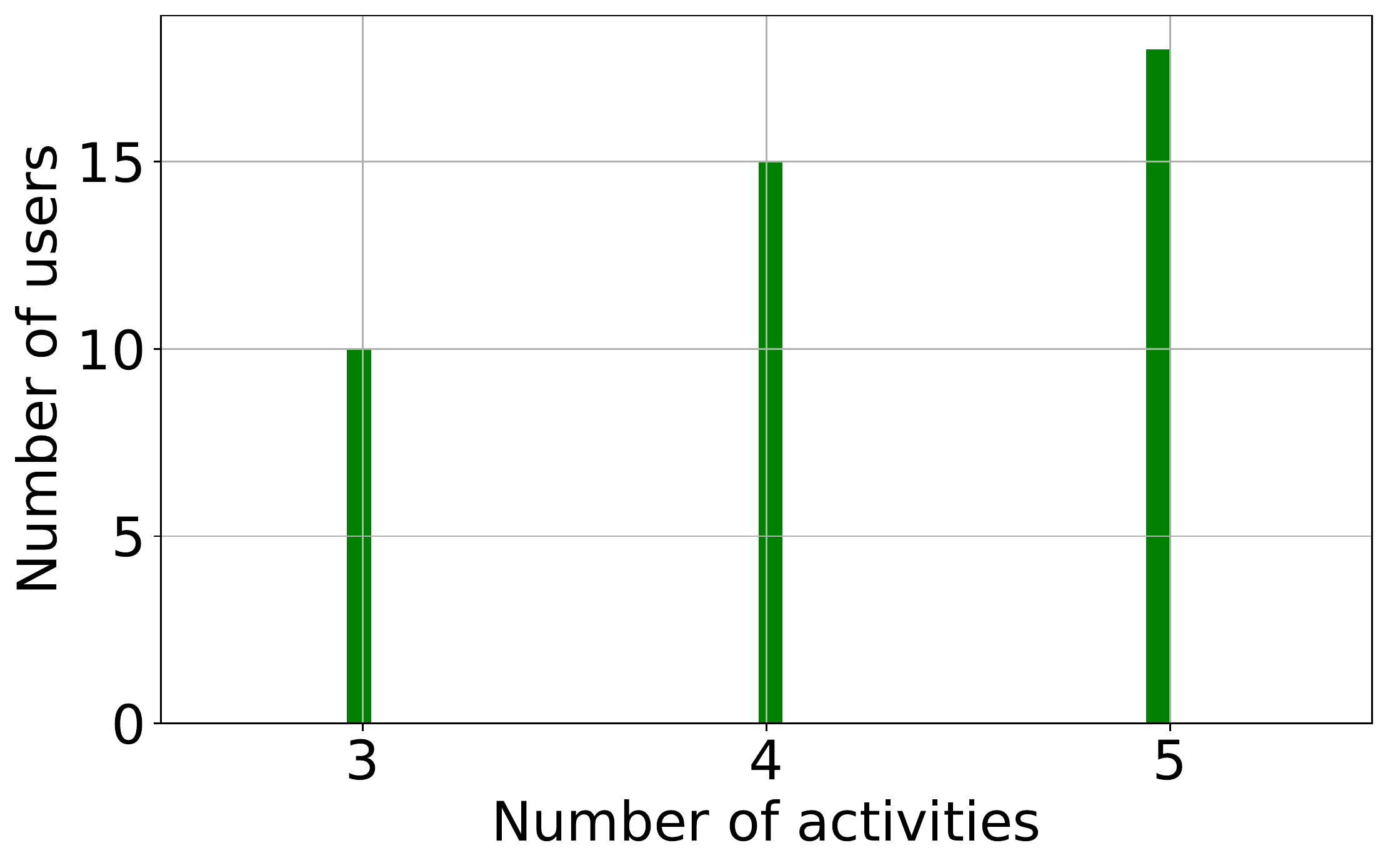}  
      \label{fig:num_hist}    
    } 
    \hspace{3mm}
    \subfigure[Sample distributions of collected dataset.]{
      \includegraphics[width=1.55in]{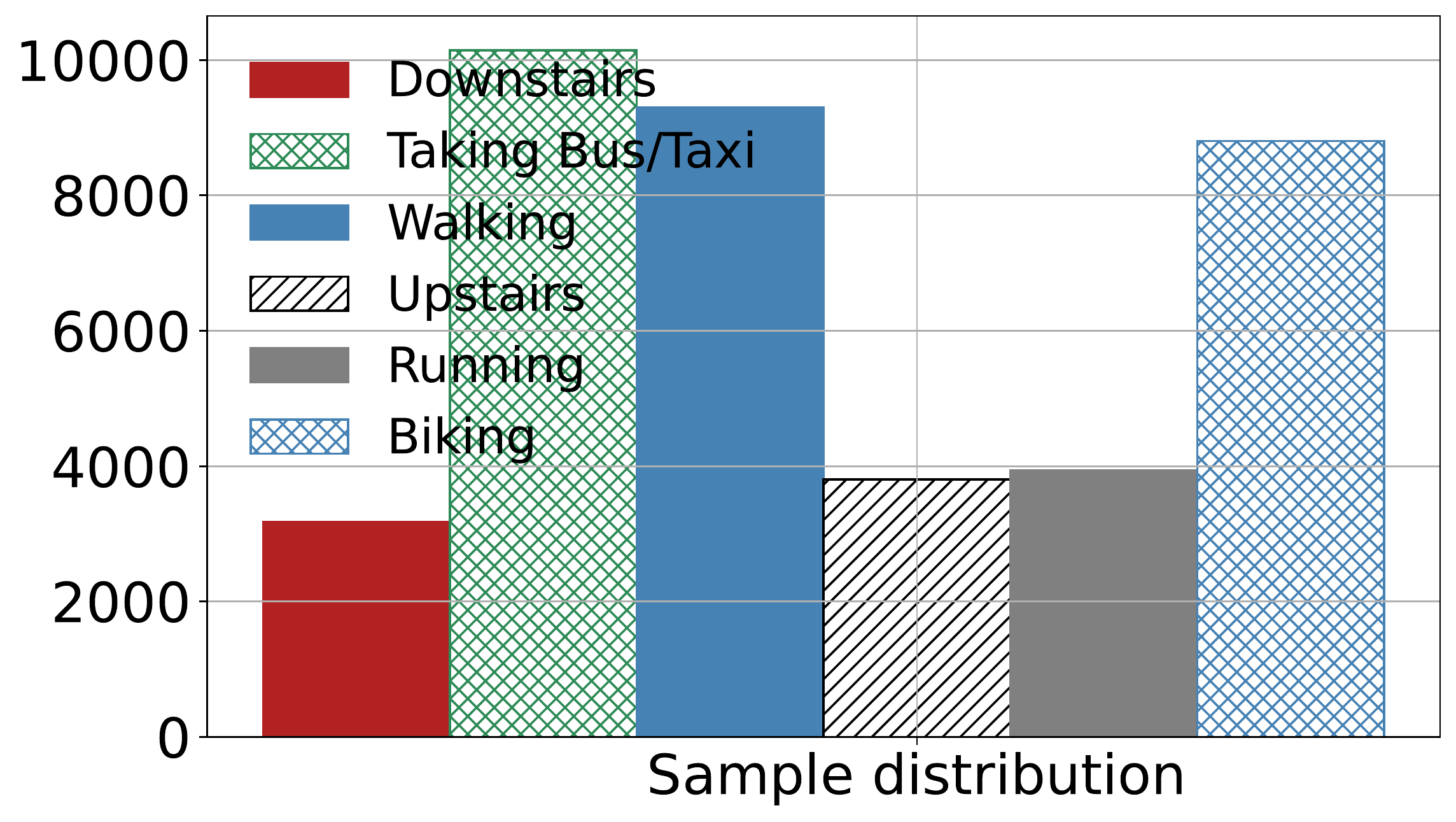}  
      \label{fig:sample_dis}
    }
  }
  \caption{(a) Distribution of samples from activity ``Stair-down'' for all the users in HHAR dataset; (b) Distributions of samples from all activities of user $h$.}
  \Description{Sample distributions.}
  \label{fig:collect_sta}
\end{figure}

For each participant, any of its activity type with uploaded ``good'' samples less than 20 will be removed. After that, We further remove users with a number of local activity types less than 3. Finally, we get 48 users each with a different number of activities as well as different activity types. 
As shown in Fig.~\ref{fig:num_hist}, after sample and user selection, we get a heterogeneity dataset with Non-IID label distributions among users. Most users have 5 types of local activity and none of them provides all the 6 types of activities we selected. Also, our collected dataset is highly biased to activities that are common in real life, e.g. ``Walking'' and ``Taking bus/taxi''. On the other hand, we can only collect a relatively small number of samples for activities that are not commonly performed in daily life, such as ``Upstairs'', ``Downstairs'' and ``Running''.

In our experiments on the collected dataset, we do not remove the unbalancedness and directly train our model on the unbalanced dataset. This is because, in a real-world federated learning setting, we can not control the dataset distribution on the user side. However, to evaluate the performance of a given model, we use a balanced test dataset, where the numbers of data samples from each activity type is balanced. 

\end{document}